\documentclass[aps,a4paper,twocolumn,prl]{revtex4-1}

\usepackage{amsmath}%
\usepackage{amsfonts}%
\usepackage{amssymb}%
\usepackage[cal=cm]{mathalfa}
\usepackage{graphicx}
\usepackage{braket}
\usepackage{sidecap}
\usepackage{color}
\usepackage{bbm}
\usepackage{hyperref}
\usepackage{nicefrac}
\usepackage{float}

\usepackage[utf8]{inputenc}

\newcommand{\cut}[1]{}
\newcommand{\mox}[2]{#2}

\begin{document}

\title{Residual error-disturbance uncertainties in successive spin-$\nicefrac{1}{2}$ measurements \\
tested in matter-wave optics}
\author{B{\"u}lent Demirel$^1$}
\author{Stephan Sponar$^1$}
\author{Georg Sulyok$^1$}
\author{ Masanao Ozawa$^2$}
\author{Yuji Hasegawa$^1$}
\email{Hasegawa@ati.ac.at}
\affiliation{%
$^1$Atominstitut, TU Wien, Stadionallee 2, 1020 Vienna, Austria \\
$^2$Graduate School of Information Science, Nagoya University, Chikusa-ku, Nagoya 464-8601, Japan }

\date{\today}

\hyphenpenalty=800\relax
\exhyphenpenalty=800\relax
\sloppy

\noindent
\begin{abstract} The indeterminacy inherent in quantum measurement is an outstanding character of quantum theory, which manifests itself typically in Heisenberg's error-disturbance uncertainty relation.
In the last decade, Heisenberg's relation has been generalized to hold for completely general quantum measurements. Nevertheless, the strength of those relations has not been clarified yet for mixed quantum states.
Recently, a new error-disturbance uncertainty relation (EDUR), stringent for generalized input states, has been introduced by one of the present authors. A neutron-optical experiment is carried out to investigate this new relation: it is tested whether error and disturbance of quantum measurements disappear or persist in mixing up the measured ensemble.
Our results exhibit that measurement error and disturbance remain constant independent of the degree of mixture. The tightness of the new EDUR is confirmed, thereby validating the theoretical prediction.
\end{abstract}

\maketitle

Quantum measurement, through which a value of a physical property is assigned, has still eluded our consistent, physical understanding \cite{ZurekWheelerBook}. Born's rule gives physical connection between the quantum-mechanical formalism and the prediction of probabilities of events occurring in a single quantum-system \cite{Born26}. Our studies, however, are not limited to
measurements of physical quantities on a single quantum-system \cite{Neumann27}, but are rather concerned with statistical ensembles of a quantum system reflecting actual circumstances. All information of physical importance is, thus, attributed to a statistical state, represented by a so-called density matrix \cite{Sakurai}. Note that there is no uniqueness of the representation of a mixed state as a convex sum of pure states \cite{Ballentine}. That is, the same mixed-state density matrix can be obtained with different \emph{blends} for that \cite{Suesmann58, Englert13}; experiments can distinguish the difference in mixture but no  evidence can be found in different generation methods of mixture. All \emph{as-if realities} consisting in blending is not accessible, turning to be virtual  \cite{Englert13}.

In this letter, we report on experimental investigations of the influence of the state mixture on error and disturbance uncertainties in successive spin-1/2 measurements. For this purpose, we generate mixed ensembles of the spin state of neutrons and tune the degree of mixture systematically.
It is well-known that the occurrence of a dephasing in double-slit experiments leads to (phase) mixture,
easily washing out interference fringes, i.e., quantum interference vanishes for mixed states and quasi-classical behavior can emerge in certain circumstances \cite{Zurek03, GiuliniBook}.
Thus, it is an interesting problem worth investigating whether the mixture of the measured ensemble increases or decreases the measurement uncertainty. 
Since all the states of a quantum system, used in practical resources, are more or less statistically mixed ensembles, our results will help to classify the practical role of a quantum effect employed in quantum technology. 

\begin{figure*}
	\includegraphics[width=.9\textwidth] {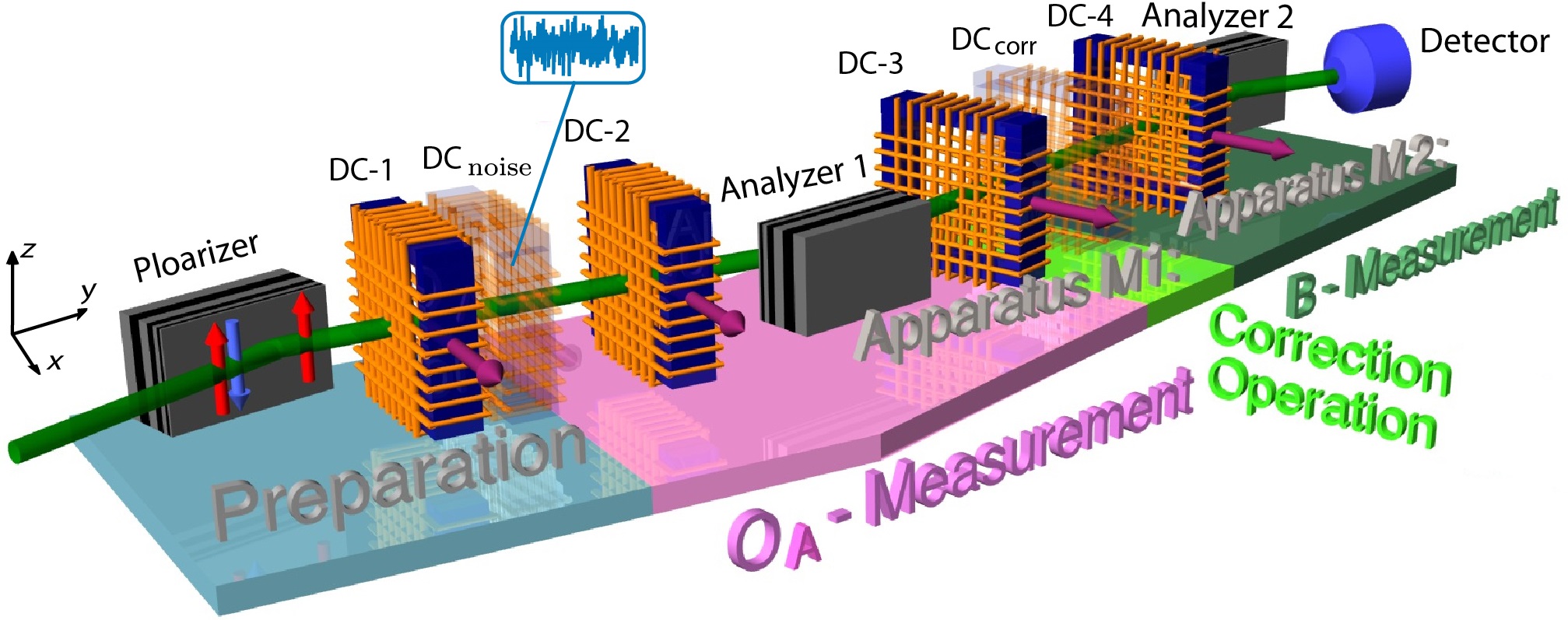}
	\caption{ Illustration of the experimental set-up. 
		The neutron-polarimeter set-up consists of three stages. 
		(1) Preparation (blue region): a monochromatic neutron beam is polarized in $+z$-direction by passing through a supermirror spin polarizer. In the coil (DC-1) the required directions of the input states are generated and the mixture is controlled via a tunable noisy magnetic field.
		(2) Apparatus M1, consisting of  a projective $O_A$ - measurement
		(pink region) and a correction operation (light green region): the first measurement is carried out by the analyzer-1 together with the coils (DC-3/4)
		followed by a unitary rotation of the output state of the  $O_A$ - measurement.
		(3) Apparatus M2, measuring $B$ (dark green region): the second measurement is fixed to make a $B$ - measurement, which is carried out by the the coil (DC-4) and the analyzer-2. Transparent coils are virtual and their positions are taken up by other DC-coils in practice.
		}
	\label{fig1:ERDURSetup}
\end{figure*}

The uncertainty principle proposed by Heisenberg \cite{Heisenberg27} in 1927 states that it is impossible to simultaneously measure two conjugate observables with arbitrary precision.
By the famous $\gamma$ ray microscope thought experiment, Heisenberg showed the error-disturbance relation $q_1p_1\sim \hbar$ for the error $q_1$ of a position measurement and the disturbance  $p_1$ thereby caused on the momentum. In his mathematical derivation of this relation, he introduced the famous preparational uncertainty relation $\Delta q \,\Delta p\ge\frac{\hbar}{2}$ for standard deviations $\Delta q$ and $\Delta p$ for position $q$ and momentum $p$; a general proof was given shortly afterward by Kennard \cite{Kennard27}. Robertson generalized this relation to an arbitrary pair of non-commuting observables $A,B$ for a given quantum state $\ket{\psi}$ replacing the lower bound $\hbar/2$ by the bound
${C}_{AB}=\frac{1}{2}\vert\bra{\psi}[A,B]\ket{\psi}\vert$
\cite{Robertson29}.

An error-disturbance uncertainty relation (EDUR) valid for an arbitrary pair of observables and for arbitrary generalized measurements was derived by Ozawa  \cite{Ozawa03, Ozawa04} as
\begin{equation}
\label{Eq:Ozawa's firstEDR}
	{\epsilon}(A){\eta}(B) + {\epsilon}(A) \Delta B + {\eta}(B)\Delta A\geq{C}_{AB} \, ,
\end{equation}
validity of which were experimentally tested with neutrons \cite{Erhart12,Sulyok13} and with photons \cite{Steinberg12, Edamatsu13}. Other approaches to measuremental uncertainty relations
can be found for example in \cite{Weston13,Busch13,Busch14,Buscemi14,Lu14}.

In pursuit of an improvement of relation~\eqref{Eq:Ozawa's firstEDR}, a stronger inequality

\begin{eqnarray}
\label{Eq:Branciard's general EDR}
{\epsilon}(A)^2 \Delta B^2 &&+ {\eta}(B)^2\Delta A^2 \nonumber\\&&+ 2{\epsilon}(A){\eta}(B)\sqrt{\Delta A^2 \Delta B^2-\mathcal{C}_{AB}^2} \geq \mathcal{C}_{AB}^2 \, 
\end{eqnarray}

\noindent
was introduced by Branciard \cite{Branciard13}. \mox{Experimental tests of this relation for pure input states were carried out by using photonic systems \cite{Ringbauer14, Kaneda14} and neutrons \cite{Sponar14}.}{}
Later, it was pointed out that the relation above is not stringent for mixed states in general, when
the Robertson bound ${C}_{AB}$ is simply extended to
$C'_{AB} = \frac{1}{2}\vert \text{Tr} ([A,B] \rho) \vert $, which decreases for mixed states and vanishes for totally mixed states \cite{Branciard14}. Further improvement of the bound was put forward by Ozawa \cite{Ozawa14} who showed that the constant $C_{AB}$ in Eq. \eqref{Eq:Branciard's general EDR} can be replaced by a stronger constant $D_{AB}$ defined by $D_{AB}= \frac{1}{2} \text{Tr}\left(|\sqrt{\rho} [A,B] \sqrt{\rho}|\right)$.
This new parameter coincides with the Robertson bound $\text{C}_{AB}$ when $\rho$ is a pure state, but makes the EDUR in the form of Eq. \eqref{Eq:Branciard's general EDR} stronger for a mixed ensemble.

All these considerations so far have been valid for a general, arbitrary pair of non-commuting observables. As the simplest case, spin-$\nicefrac{1}{2}$ observables, represented by a set of Pauli operators, have been a major focus of investigations of EDURs. Branciard
\cite{Branciard13} showed that
for binary measurements with $A^2 = B^2 = \mathbbm{1}$ and $\braket{A} = \braket{B} = 0$,
where $\braket{\cdots}$ stands for the expectation value in the system state,
 Eq. \eqref{Eq:Branciard's general EDR} can be strengthened to
a stronger EDUR.
Ozawa demonstrated that it can be further strengthened by replacing again the bound $C'_{AB}$
by $D_{AB}$ for mixed spin states \cite{Ozawa14}.



\mox{}{
We carried out a neutron polarimeter experiment at the research reactor in Vienna as depicted in FIG. \ref{fig1:ERDURSetup} . The incident neutrons with a wavelength $\lambda \cong 2.02$ \AA,
are polarized by the first supermirror. A guide field between Polarizer/Analyzer 1 and between Analyzer1/Analyzer 2 in +\textit{z}-direction is applied and determines the quantization axis. Building on the previous performance of the studies of the EDUR for pure states \cite{Erhart12,Sulyok13,Sponar14}, we extend here the investigation by applying two procedures, i.e. the generation of mixed states and modification of the first measurement process in apparatus (M1) by unitary transforming the output states. The former allows the study of the EDUR for mixed states and the latter enables to tune the disturbance.
}

The polarimeter set-up consists of \mox{four}{three} stages \mox{just like  in FIG. \ref{fig1:ERDURSetup}}{}: (1) the state preparation, (2) the apparatus M1  performing a projective $O_A$ - measurement plus the correction procedure and (3) the apparatus M2 performing a  $B$ - measurement. \mox{The DC-coils in the set-up generate magnetic fields in the \textit{x}-direction and induce a rotation of the neutron spin, commonly referred to as Larmor precession, which}{Larmor precession induced by magnetic fields $B_x$ in the DC-coils} allows to orient all required directions of the spin-measurements. 
The mixing of the state can be tuned by \mox{applying}{} a noise magnetic field $B_{\text{noise}}$ \cite{Klepp08}.
In practice we realize $\pi/2$-rotations with noisy fields by one DC-coil
(DC-1), where the required mixture can be adjusted by the amplitude of the noise signal.

In stage one, the input states were chosen to be $\rho_x(\alpha)= \frac{1}{2}(\mathbbm{1}+\alpha \sigma_x)$ with five different mixtures $\alpha = \{1, 0.75, 0.5, 0.25, 0 \}$. The degree of mixture was verified by measuring the expectation values of the Pauli-spin operators $\text{Tr}(\sigma_i \rho_x)$ for $i={x,y,z}$ each.
Typical fidelity $F = \text{Tr} \left(\sqrt{ \sqrt{\rho_x} \rho^{\rm exp}_x \sqrt{\rho_x}}\right)$ of the pure input state $\rho_x$ was 0.982(5).
The so-called `three-state-method' \cite{Ozawa04} is applied to acquire the values of $\epsilon(A)$ and $\eta(B)$ \mox{: the three different states $\rho, A \rho A$ and $\rho_{|A} = P^{+}_{A} \rho P^{+}_{A}/\text{Tr}(P^{+}_{A} \rho)$ are sent to the apparatuses. Here $P^{+}_{A} = \frac{1}{2}(\mathbbm{1}+A)$ is the eigen-projection of the positive eigenvalue of observable A. Error $\epsilon(A)$ and disturbance $\eta(B)$ are determined in the same manner}{} (see Supplement for more details).

The second stage represents the apparatus M1
in which the coil DC-2/3 plus Analyzer 1 perform the projective measurement that actually measures the observable $O_{A} = \text{cos}(\theta_{OA}) \sigma_z + \text{sin}(\theta_{OA}) \sigma_y$;  $\theta_{OA}$ is the detuning angle of this measurement and leaves the neutron in the $\ket{O_A=\pm1}$ states.
In the correction stage,
this output sate of the $O_A$ - measurement is transformed by a unitary operator $U^{\rm corr}$.
One can realize the optimal (and anti-optimal) correction by adjusting $U^{\rm corr}$. Note that in our previous study \cite{Erhart12,Sulyok13,Sponar14} the unitary operation $U^{\rm corr}$
is not \mox{used}{applied} and fixed as $U^{\rm corr}=\mathbbm{1}$ in practice.
The last stage consists of the successive measurement of $B = \sigma_y$ in apparatus M2 which is accomplished again by a DC-coil (DC-4) plus analyzer (Analyzer 2) combination. \mox{}{Note that the final spin rotation is not applied, since it has no influence on measured intensities. }

We investigate a neutron spin measurement in which $A = \sigma_z$, $B = \sigma_y$ and consider a general mixed ensemble represented by $\rho = \frac{1}{2}(\mathbbm{1}+\textbf{r} \cdot \boldsymbol{\sigma})$ satisfying $\braket{A} = \braket{B} = 0$; then, $\rho$ is generally parameterized as
$\rho_x(\alpha) = \frac{1}{2}(\mathbbm{1}+\alpha \sigma_x)$.
In this case, the parameter $D_{AB}=1$ is constant and yields the tight relation
\cite{Ozawa14}
\begin{equation}
\label{eq:CircularQbitEDR}
\left(\epsilon(A)^2-2\right)^2 + \left(\eta(B)^2-2\right)^2\leq 4 \,
\end{equation}
\noindent
for any mixed states $\rho_x(\alpha)$, while
the parameter ${C}'_{AB}$ depends on the mixture, i.e. the length $|\textbf{r}|$ of the vector.
Experimental tests of this relation for pure input states were carried out
by using photonic systems \cite{Ringbauer14, Kaneda14} and neutrons \cite{Sponar14}.

\begin{figure}
	\includegraphics[width=.5\textwidth]{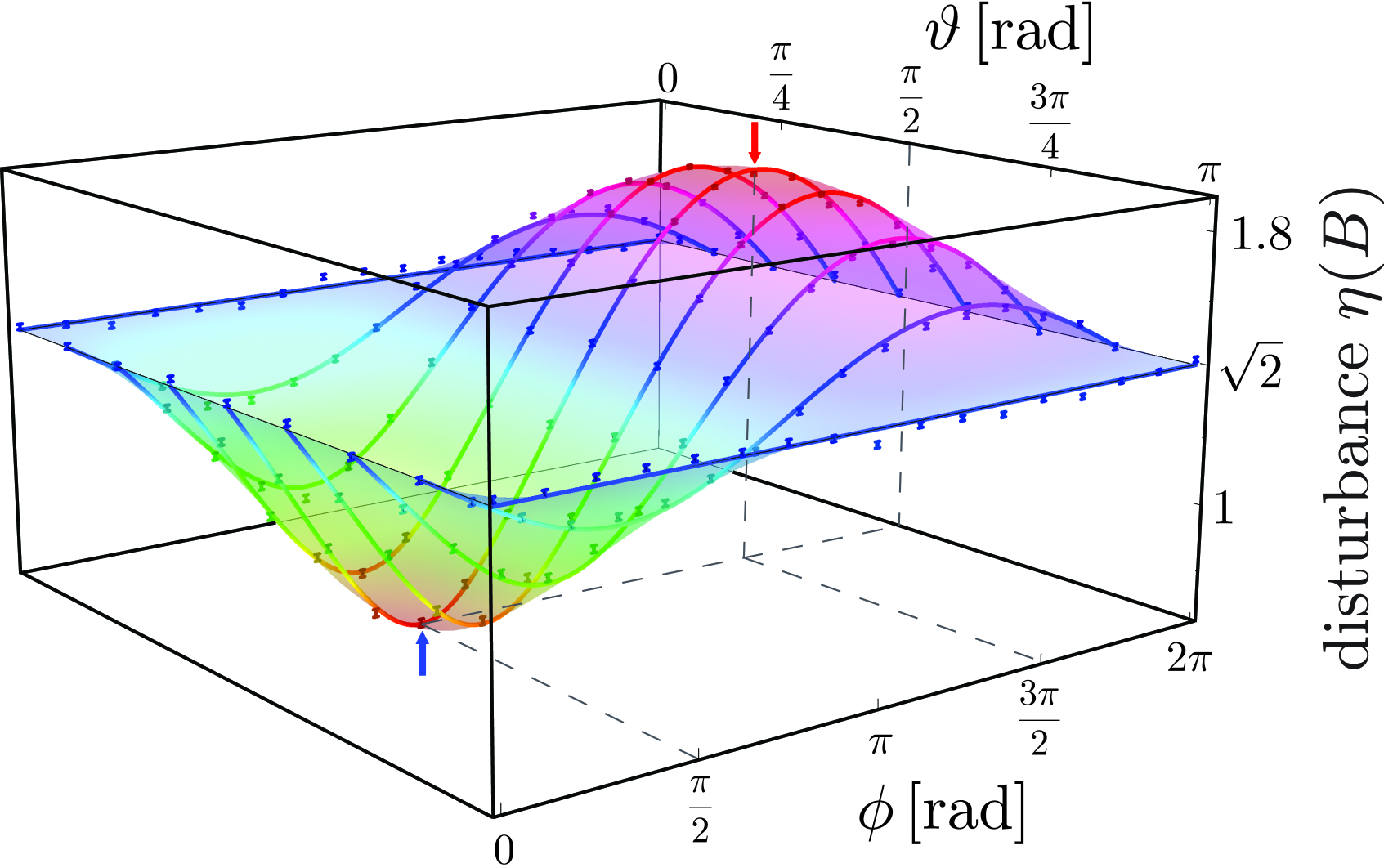}
	\caption{Influence of the correction procedure on the disturbance.
		After the projective measurement of $O_A \left(\theta_{OA} = 5 \pi /18 \right)$
		plus unitary rotations $U^{\rm corr}$ (with angle parameters $(\vartheta, \phi)$ for the output state
		of the apparatus M1),
		the measurement of $B=\sigma_y$ is performed in apparatus M2.
		The angles identify the output states $\ket{ \psi (\vartheta, \phi)} = \left(\text{cos}(\vartheta /2), e^{i \phi} \text{sin}(\vartheta /2)\right)^T$ and $\ket{- \psi (\vartheta, \phi)} = \ket{\psi (\pi-\vartheta, \phi+\pi)}$ of the unitary operation. Blue and red arrow indicate the position of the minimal
		\mox{($\pi/2,\pi/2$) and maximal ($\pi/2,3\pi/2$)}{($\frac{\pi}{2}$,$\frac{3\pi}{2}$) and maximal ($\frac{\pi}{2}$,$\frac{\pi}{2}$)}
		disturbance.}
	\label{fig:Correction_Search}
\end{figure}

\begin{figure}
	\includegraphics[width=0.44\textwidth]{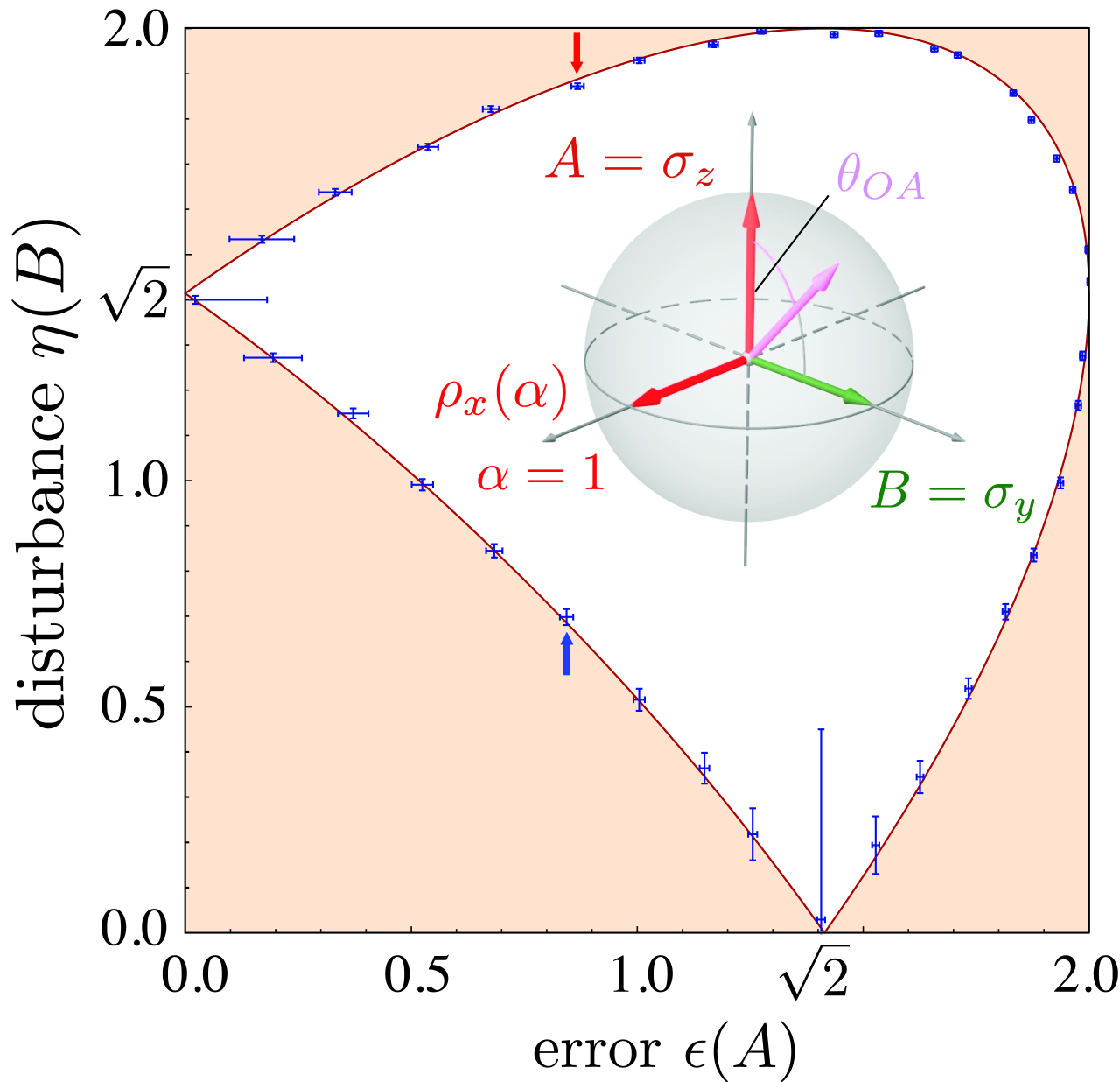}
	\caption{Error-disturbance uncertainty relation as indicated by inequality Eq. \eqref{eq:CircularQbitEDR} measured with pure states: not only the lower but also upper bounds of the disturbance (at fixed error) are found. For a detuning angle of $\theta_{OA}=0$ the output observable $O_A$ coincides with $A = \sigma_z$ at which point $(\epsilon(A), \eta(B)) = (0, \sqrt{2})$. For increasing angles $\theta_{OA}$ the error increases as well and disturbance spreads between the maximum and minimum values. The extremal points are reached at $\theta_{OA}=\mox{\pi/2}{\frac{\pi}{2}}$, at which $O_A$ equals $B$. For angles from $\mox{\pi/2}{\frac{\pi}{2}}$ to $\pi$ the EDUR evolves back. Blue and red arrows indicate the points denoted to the maximal and minimal disturbance in FIG. \ref{fig:Correction_Search}.}
	\label{fig:Circular_ErrorDisturbance}
\end{figure}

\begin{figure*}
	\includegraphics[width=1.0\textwidth]{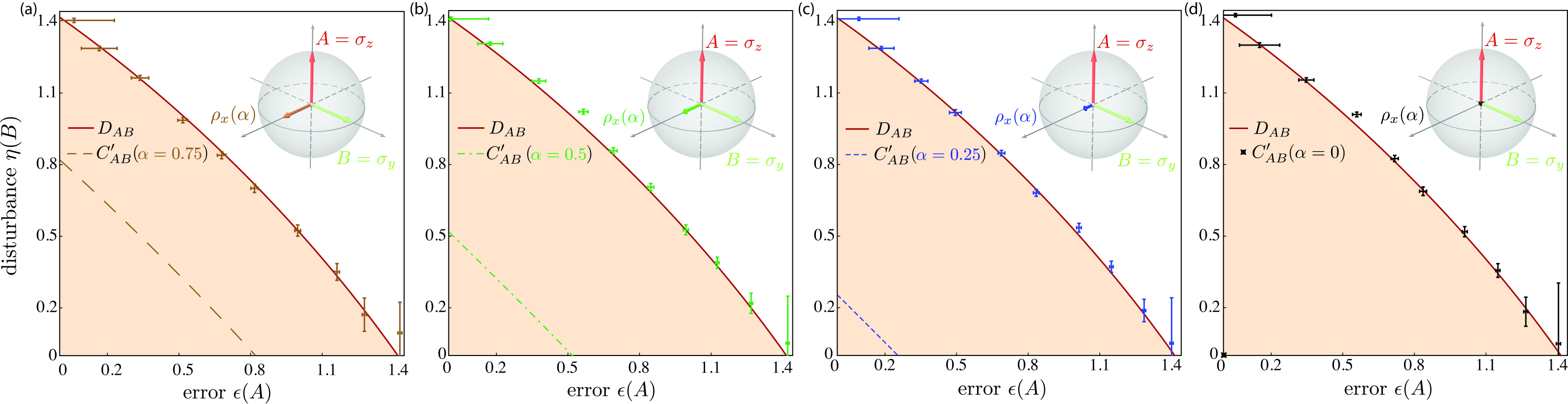}
	\caption{Error $\epsilon(A)$ versus disturbance $\eta(B)$ for the standard configuration $(A = \sigma_z, B = \sigma_y)$ with four different mixtures of the state $\rho_x (\alpha) = \frac{1}{2}(\mathbbm{1}+\alpha \sigma_x)$: (a) $\alpha=0.75$, (b) $\alpha=0.5$, (c) $\alpha=0.25$ and (d) $\alpha=0$.  The red shaded areas are forbidden according to Eq.~\eqref{eq:CircularQbitEDR}. The border indicates the lower bound $D_{AB} = 1$ which is saturated by our data points. The theoretical behavior  of the bound $C'_{AB}$ is indicated by the colored dashed lines. A change of the mixture parameter $\alpha$ has no effect on the final error-disturbance relation in the standard configuration as initially predicted by the expectation value $C'_{AB}$.}
	\label{fig:RhoX_results}
\end{figure*}

\begin{figure}
	\includegraphics[width=.37\textwidth]{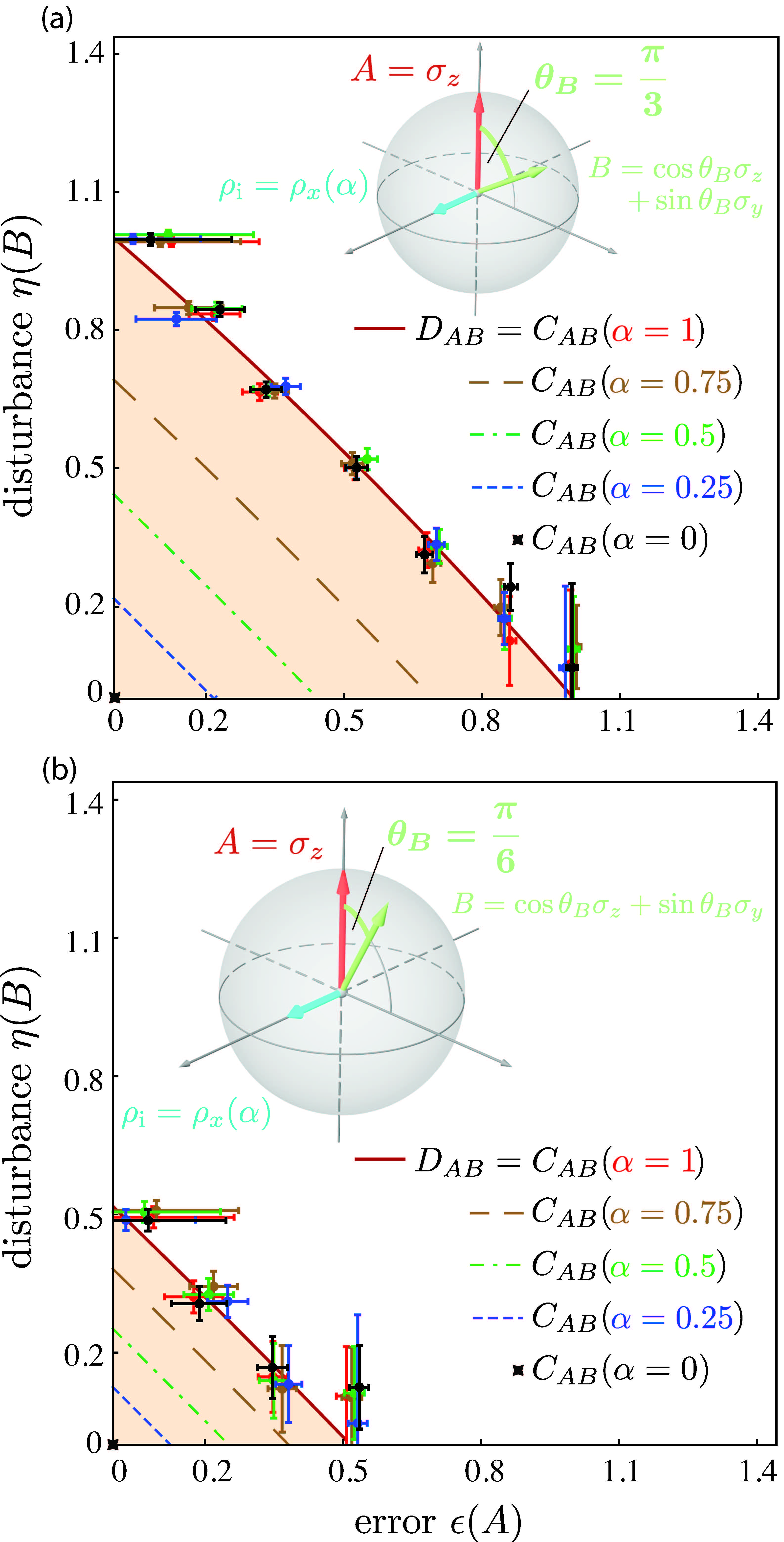}
	\caption{Results of error and disturbance for a different observable $B$, parametrized by the angle $\theta_B$. The bound depends on the 'degree of commutativity' $[A,B]$ but not on the input states. Analogous to FIG. \ref{fig:RhoX_results} the results were recorded with five different mixtures. (a) Result for  $\theta_B = \pi/3$. (b) Result for  $\theta_B = \pi/6$.}
	\label{fig:RhoX_results_B60_30}
\end{figure}

Our first study examines the influence of the unitary transformation in apparatus (M1). First, pure input states are generated and the detuning angle $\theta_{OA}$ is fixed at $5 \pi /18$. Then, the eigenstate of $O_A$ after apparatus M1 is unitarily transformed  to the state $\ket{\psi (\vartheta, \phi)}$, given by $\ket{ \psi (\vartheta, \phi)} = \left(\text{cos}(\vartheta /2), e^{i \phi} \text{sin}(\vartheta /2)\right)^T$.
The measured disturbance as a function of the polar and azimuthal angle $(\theta, \phi)$ is plotted in FIG. \ref{fig:Correction_Search}.
This plot clearly exhibits the reduction and the enhancement of disturbance by the choice of $\vartheta$ and $\phi$.
In addition, it is shown that the minimal and maximal disturbances, illustrated by blue and red arrows in FIG. \ref{fig:Correction_Search}, are achieved when the state after measurement is unitarily transformed into eigenstates of the observable $B = \sigma_y$.
(see Supplement for theoretical details of the correction/ anti-correction procedure).

After determination of maximum and minimum values of the disturbance,
the EDUR given by Eq. \eqref{eq:CircularQbitEDR} is analyzed. The experimentally determined error versus maximum and minimum disturbances are plotted in FIG. \ref{fig:Circular_ErrorDisturbance} for pure states together with the theoretically predicted bound. The red shaded area marks the forbidden region. The lower and upper bound was measured for angle $\theta_{OA}=[0,\pi]$ with a step of $\pi/18$. For $\theta_{OA} = 0$ we have $\epsilon(A) = 0$ at which point the disturbance is unique. When $\theta_{OA}=\pi/2$ $(O_A=B)$,
the disturbance reaches it's [maximum] minimum value, depending on the unitary [anti-] optimal correction transformation. When $\theta_{OA} = \pi$, $O_A = - A$ the error is maximal and disturbance is independent of the transformation once again. 

Next the influence of the mixture of the input states is studied. By applying the optimal correction procedure the minimal disturbances are measured tuning the input states $\rho_x (\alpha) = \frac{1}{2}(\mathbbm{1}+\alpha \sigma_x)$. The results are plotted in  FIG. \ref{fig:RhoX_results}. Each plot exhibits optimal EDUR for a particular mixture with theoretical predictions by $D_{AB}$ and $C'_{AB}$. It is immediately seen that the error-disturbance uncertainty is insensitive to dephasing or amplitude damping of the input states  caused by the fluctuating magnetic field and that the bound is preserved perfectly.
The measured values always saturate inequality Eq. \eqref{eq:CircularQbitEDR}, for mixed spin states no dependence on mixture appears. Only the bound given by $D_{AB}$ leads to saturation of the error-disturbance uncertainty relation. This statement is  also true for different choices of the observables $A$ and $B$.
If an extended configuration including non-maximally incompatible pairs of observables $A$ and $B$ is considered, e.g. $B = \text{cos}(\theta_B)\sigma_z+\text{sin}(\theta_B)\sigma_y$ then minimal disturbance is given by $\eta(B) =2\, \text{sin} \left|  (\theta_{OA}-\theta_{B})/2  \right|$. 	The results for two different angles $\theta_B = \pi/3$ and $\theta_B = \pi/6$ are plotted in  FIG. \ref{fig:RhoX_results_B60_30}. For pure input-states, according to the change of the commutator $[A,B]$, both parameters $C'_{AB}$ and $D_{AB}$ represent the lower bound of the EDUR. 
For mixed input-sates, however, only the bound $D_{AB}$ explains the correct behavior.

The successive nature of the measurement made it obvious how the correction procedure, i.e., a unitary transformation, can be incorporated to the whole measurement. Disturbance is strongly affected by this correction and we have observed the maximum and the minimum disturbance by optimal and anti-optimal corrections.
Our experiment successfully demonstrates the tightness of the bound $D_{AB}$ and the non-tightness of the simply extended Robertson bound $C'_{AB}$.
We confirmed the independence of the EDUR on the mixture of the states for the case of dichotomic observables $A$, $B$ with $\braket{A} = \braket{B} = 0$.
This is considered to be due to the fact that the observed uncertainty for Pauli operators is originated more in observables than in input states: this reminds us another state-independence appearing in quantum contextuality, which was confirmed in an ion experiment \cite{Kirchmair2009}.
Since quantum states, practically used in application such as quantum communication and computation, are more mixed ensemble due to (unavoidable) dephasing and decoherence than in a laboratory, our study shed a light on the new aspects of quantum measurements available for practical applications.

We acknowledge support by the Austrian Science Fund, FWF (Nos. P27666-N20 and P24973-
N20), the European Research
Council, ERC (No. MP1006), the John Templeton Foundations (ID 35771) and  
Japan Society for the Promotion of Science, JSPS KAKENHI (No. 26247016).
We thank H. Rauch, (Vienna), M.J.W. Hall (Brisbane), F. Busemi (Nagoya) and
A. Hosoya (Tokyo) for their helpful comments.



%

\newpage

\onecolumngrid

\appendix

\renewcommand{\theequation}{S\arabic{equation}}
\setcounter{equation}{0}  
 
\renewcommand{\thefigure}{S\arabic{figure}}
\setcounter{figure}{0}  
  
\section*{Supplementary Material}  


\section{Theoretical framework}
\label{sec:theory}

\textbf{Theory of error and disturbance.}
Any apparatus \textbf{M} is described by an indirect measurement model $
(\mathcal{K}, \ket{\xi}, U, M)$, where $\mathcal{K}$ is the apparatus state space,
$\ket{\xi}$ is the initial apparatus state, $U$ is the unitary operator describing the
object-apparatus interaction, and $M$ is the meter observable of the apparatus \cite{Ozawa04B}.
Let $\rho$ be
the initial object state, the error $\epsilon(A)$ for measuring an observable $A$
and the disturbance $\eta(B)$ caused on an observable $B$
are defined as \cite{Ozawa03B}
\begin{equation}
	\label{GeneralDefintionOfEpsEta}
	\begin{split}
		\epsilon(A)^2\! =\! \text{Tr}\left[\left(U^{\dagger}(\mathbbm{1} \otimes M)U-A\otimes \mathbbm{1}\right)^2 \rho \otimes \ket{\xi}\!\bra{\xi}\right], \\
		\eta(B)^2 \!=\! \text{Tr}\left[  \left(U^{\dagger}(B \otimes \mathbbm{1})U-B\otimes \mathbbm{1}\right)^2 \rho \otimes \ket{\xi}\!\bra{\xi}\right].
	\end{split}
\end{equation}
\noindent
To further evaluate error and disturbance,
we suppose that the meter observable $M$ has non-degenerate eigenvalues $m$ with spectral decomposition $M = \sum m \ket{m} \bra{m}$. Then, the apparatus \textbf{M} is
characterized by the family of measurement operators \{$M_{m}$\}  defined by $M_{m} = \braket{m|U|\xi}$.
The positive operator-valued measure (POVM) of \textbf{M} is the family $\{P_m \}$
of positive operators defined by $P_m = M^{\dagger}_{m}M_{m}$.
We can rewrite error and disturbance, assuming $P_{m}$ are mutually orthogonal projectors, as
\begin{equation}
	\begin{split}
		\epsilon(A)^2 &=  \braket{(O_A -A)^2}+ \braket{(O_A^{(2)} -O_A^2)} \, ,\\
		\eta(B)^2 &= \braket{(O_B -B)^2}+ \braket{(O_B^{(2)} -O_B^2)}\, ,
	\end{split}
\end{equation}
\noindent
where the output operators are given by $O^{(k)}_A = \sum_m m^k P_m$ and $O^{(k)}_B = \sum_m {M_m}^{\dagger} B^k {M_m}$.
As usual, we require the meter observable $M$ to have the same spectrum as the measured observable $A$.  For binary observables $A^2=B^2=\mathbbm{1}$, we have
$O_A^{(2)}=O_B^{(2)}=\mathbbm{1}$, and we obtain
\begin{equation}\label{SpecialDefintionOfEpsEta}
	\epsilon(A)^2=2-2\Re\braket{AO_A},\quad\eta(B)^2=2-2\Re\braket{B O_B}.
\end{equation}

\textbf{Optimization of disturbance.}
The measurement operators of the projective measurement of
$O_{A} = \text{cos}(\theta_{OA}) \sigma_z + \text{sin}(\theta_{OA}) \sigma_y$
carried out by
coil DC-2/3 plus Analyzer 1 are given by
$\{M_m\}=\{\ket{m_{O_A} }\bra{m_{O_A} }\}$,
where $\ket{m_{O_A}}=\ket{O_A=m}$ for $m=\pm 1$.
The coil DC$_{\rm corr}$ accounts for the unitary transformation $U^{\rm corr}$
after the projective $O_A$ - measurement and before the $B$ - measurement,
which modifies the output state of the projective $O_A$ - measurement
to attain the optimal or anti-optimal bounds for the disturbance
as suggested in \cite{Branciard13B} in the pure state case.
Thus, the measurement operators of apparatus M1 are modified as
$\{M_{m}\} = \{U^{\rm corr}\ket{m_{O_A}}\bra{m_{O_A} }\}$
without changing the POVM $\{P_m\}=\{\ket{m_{O_A}}\bra{m_{O_A} }\}$
and the output operator
$O_A=\sum_m mP_m$.

\noindent
From Eq.~\eqref{SpecialDefintionOfEpsEta} apparatus M1 has the error $\epsilon (A) = 2\, \text{sin}\frac{\theta_{OA}}{2}$. For the calculation of the disturbance $\eta(B)$ we consider the following observable: $B=\text{cos}(\theta_B)\sigma_z+\text{sin}(\theta_B)\sigma_y$, where $0\le (\theta_B)\le\frac{\pi}{2}$. The angel $\theta_B$ quantifies the closeness of the observables $A$ and $B$, where maximal incompatibility is attained for the angle $\theta_B = \frac{\pi}{2}$. In this case we get $D_{AB}=\sin\theta_{B}$, optimal and anti-optimal corrections are carried out, minimal and maximal
disturbance is given by

\begin{equation}\label{optima}
	2\, \left|\sin \frac{\theta_{OA}-\theta_{B}}{2}\right|
	\le \eta(B) \le
	2\,  \cos\frac{\theta_{OA}-\theta_{B}}{2}.
\end{equation}
To show the above, let
$B_{m}=\braket{m_{OA}|U^{\rm corr}{}^{\dagger}BU^{\rm corr}|m_{OA}}$
for $m=\pm1$.
Then we have

\begin{equation}
	\begin{gathered}
		O_B=  \sum_{m= \pm 1} \ket{m_{O_A} }B_{m}\bra{m_{O_A}}
		= \frac{(B_{+}+B_{-})}{2}\mathbbm{1}+\frac{(B_{+}-B_{-})}{2}O_{A},\\
		\Re\braket{ B O_B} = \frac{(B_{+}-B_{-})}{2}\cos(\theta_{B}-\theta_{OA}).\\
	\end{gathered}
\end{equation}

Since the extreme values are given by $\frac{(B_{+}-B_{-})}{2}=\pm 1$,
the optimal and anti-optimal values of $\eta(B)$ are given by
\begin{equation}
	\eta(B)^2 = 2 \mp 2 \Re{\braket{B O_B}} = 2 \mp 2\, \cos(\theta_{B}-\theta_{OA}).
\end{equation}
Consequently
\begin{equation}
	4\sin^2\frac{\theta_{B}-\theta_{OA}}{2}\le \eta(B)^2
	\le 4\cos^2\frac{\theta_{B}-\theta_{OA}}{2},
\end{equation}
and Eq.~\eqref{optima} follows.

\begin{figure}
	\centering
	\includegraphics[width=0.95\textwidth]{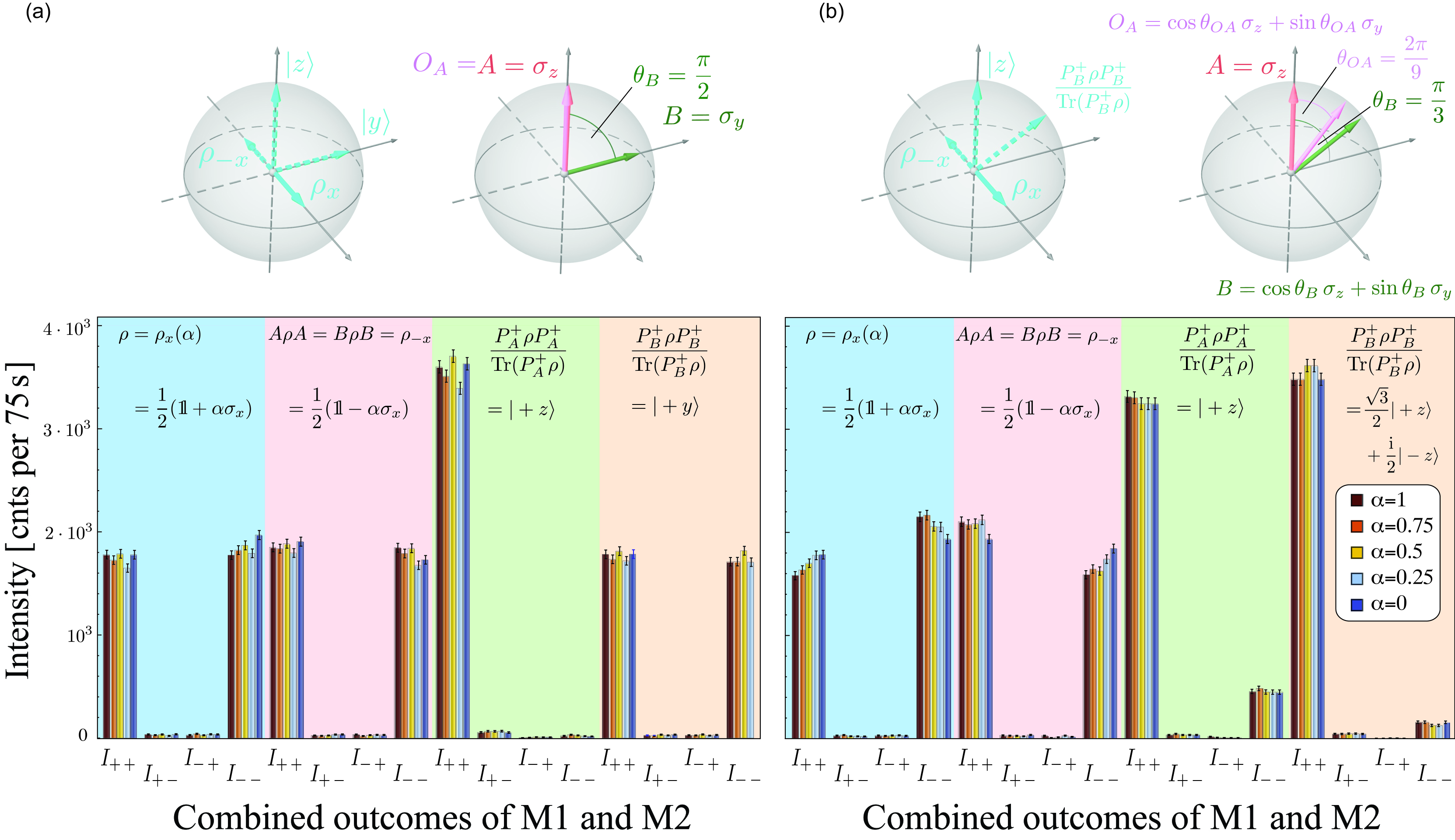}
	\caption{Detected count rates of the successive measurements carried out by apparatuses M1 and M2. The successive measurements of $O_A$ and $B$ have four outcomes, denoted as $I_{+\,+}\,\,\,I_{+\,-}\,\,\,I_{-\,+}\,\,\,I_{-\,-}$, which are  recorded for four input states, i.e.  $\rho, A \rho A=B \rho B$, $\rho_{|A} = P^{+}_{A} \rho P^{+}_{A}$ and $\rho_{|B} = P^{+}_{B} \rho P^{+}_{B}$. The initial states are given by $\rho_x(\alpha) =\frac{1}{2}(\mathbbm{1}+\alpha \sigma_x)$ with five different mixtures $\alpha = \{1, 0.75, 0.5, 0.25, 0 \}$. The observable $A$ is set as
		$\sigma_z$ and the polar angle of the observable $B$ is tuned as $\theta_B=\frac{\pi}{2}$ (for the standard configuration) and $\frac{\pi}{3}$.
		The detuned observable $O_A$ are adjusted within the $zy$-plane with polar angles $(\theta_{OA}, \theta_{B})$ given by $(0, \frac{\pi}{2})$ in (a) and $(\frac{2\pi}{9}, \frac{\pi}{3})$ in (b). Error bars represent one standard deviation of the normalized intensities. Some error bars are at the size of the markers.}
	\label{fig:fqd}
\end{figure}

\section{Data Treatment}
\label{sec:data}

\textbf{Three state method.} A re-ordering of the previous expressions \eqref{SpecialDefintionOfEpsEta} of error and disturbance allows one to obtain them by measuring the mean values of $O_A$ and $O_B$ in three different states.
We have
\begin{equation}
	\label{ExperimentalErrorDisturbance}
	\begin{split}
		\epsilon(A)^2
		=&2 - 4\,\text{Tr}(P^{+}_{A} \rho)\, \text{Tr}( \rho_{|A} O_A)+ \text{Tr}(A \rho A \, O_A)
		+ \text{Tr}(\rho O_A)\, .\\
		\eta(B)^2=&2- 4\text{Tr}(P^{+}_{B} \rho)\, \text{Tr}( \rho_{|B} O_B)+ \text{Tr}(B \rho B \, O_B)
		+ \text{Tr}(\rho O_B) \, .
	\end{split}
\end{equation}

For each projector combination of $O_A$ and $B$ an intensity output is acquired and the expectation value is calculated by combination of four intensities. We label these intensities as $I_{m b}$ where $m,b$ take values $\pm 1$. The expectation value of $O_A$ and $B$ for any state $\rho$ are obtained by
\begin{equation} \label{TraceOutputOperator}
	\text{Tr} (O_A \rho) = \frac{\sum_{m,b} m I_{mb}}{ \sum_{m,b} I_{mb}} \,, \qquad \text{Tr} (O_B \rho) = \frac{\sum_{m,b} b I_{mb}}{ \sum_{m,b} I_{mb}} \, .
\end{equation}

To determine the error $\epsilon(A)$ these intensities have to be measured for the state $\rho$, the reflected state $A \rho A$ and the pure state $\rho_{|A} = P^{+}_{A} \rho P^{+}_{A}/\text{Tr}(P^{+}_{A} \rho)$. The same applies to the measurement of disturbance $\eta(B)$ where the input states are  $\rho$, $B \rho B$ and the pure state $\rho_{|B} = P^{+}_{B} \rho P^{+}_{B}/\text{Tr}(P^{+}_{B} \rho)$. The prefactors $\text{Tr}(P^{A}_{+} \rho)$ and $\text{Tr}(P^{B}_{+} \rho)$ in Eq. \ref{ExperimentalErrorDisturbance} are obtained separately in the state preparation adjustment process.
If $\rho = \frac{1}{2}(\mathbbm{1}+ \sum_j\, r_j \sigma_j)$ is a general mixed qubit state then the polarization of the state is given by $\text{Tr}(\sigma_i \rho) = r_i$.
This relation allows to prepare and check the initial state's
direction and the degree of mixtures.
\newline

\textbf{Experimental determination of error and disturbance.} Here, we consider an extended configuration including non-maximally incompatible pairs of
observables $A$ and $B$.
The observable $A$ is left as $A=\sigma_z$ but $B$ is set as $B=\text{cos}(\theta_B)\sigma_z+\text{sin}(\theta_B)\sigma_y$, where $0\le (\theta_B)\le\frac{\pi}{2}$.
The measuring apparatus M1 performs a projective spin measurement of the
observable $O_A=\cos(\theta_{OA})\sigma_z+\sin(\theta_{OA})\sigma_y$,
where $0\le \theta_{OA}\le\frac{\pi}{2}$, followed by
the unitary operation $U^{\rm corr}$ as in the main text, and the measurement operators
of apparatus M1 are given by $\{M_m\} = \{U^{\rm corr}\ket{m_{O_A}}\bra{m_{O_A}}\}$.
Furthermore, apparatus M2 carries out the projective measurement of $B$ immediately
after the measurement carried out by apparatus M1.
Let $O_B = \sum_m {M_m}^{\dagger} B{M_m}$.
Then, the error $\epsilon(A)$ and the disturbance $\eta(B)$ are given by Eq. \eqref{optima}.
By the 'three state method', the error $\epsilon(A)$ and the disturbance $\eta(B)$
can be experimentally obtained as a sum of expectation values of the outputs from apparatus M1
and M2 in three different state as in  Eq. \eqref{ExperimentalErrorDisturbance}.
For the determination of error $\epsilon(A)$ and disturbance $\eta(B)$,
the expectation values of $O_A$ and $O_B$ in a state $\rho$ in Eq. \eqref{ExperimentalErrorDisturbance}
are derived from the intensities of the four possible outputs of the measurement $O_A$ and $B$ denoted as $I_{+\,+}\,\,\,I_{+\,-}\,\,\,I_{-\,+}$ and $I_{-\,-}$ as given by Eq. \eqref{TraceOutputOperator}.

\begin{figure}
	\centering
	\includegraphics[width=0.5\textwidth]{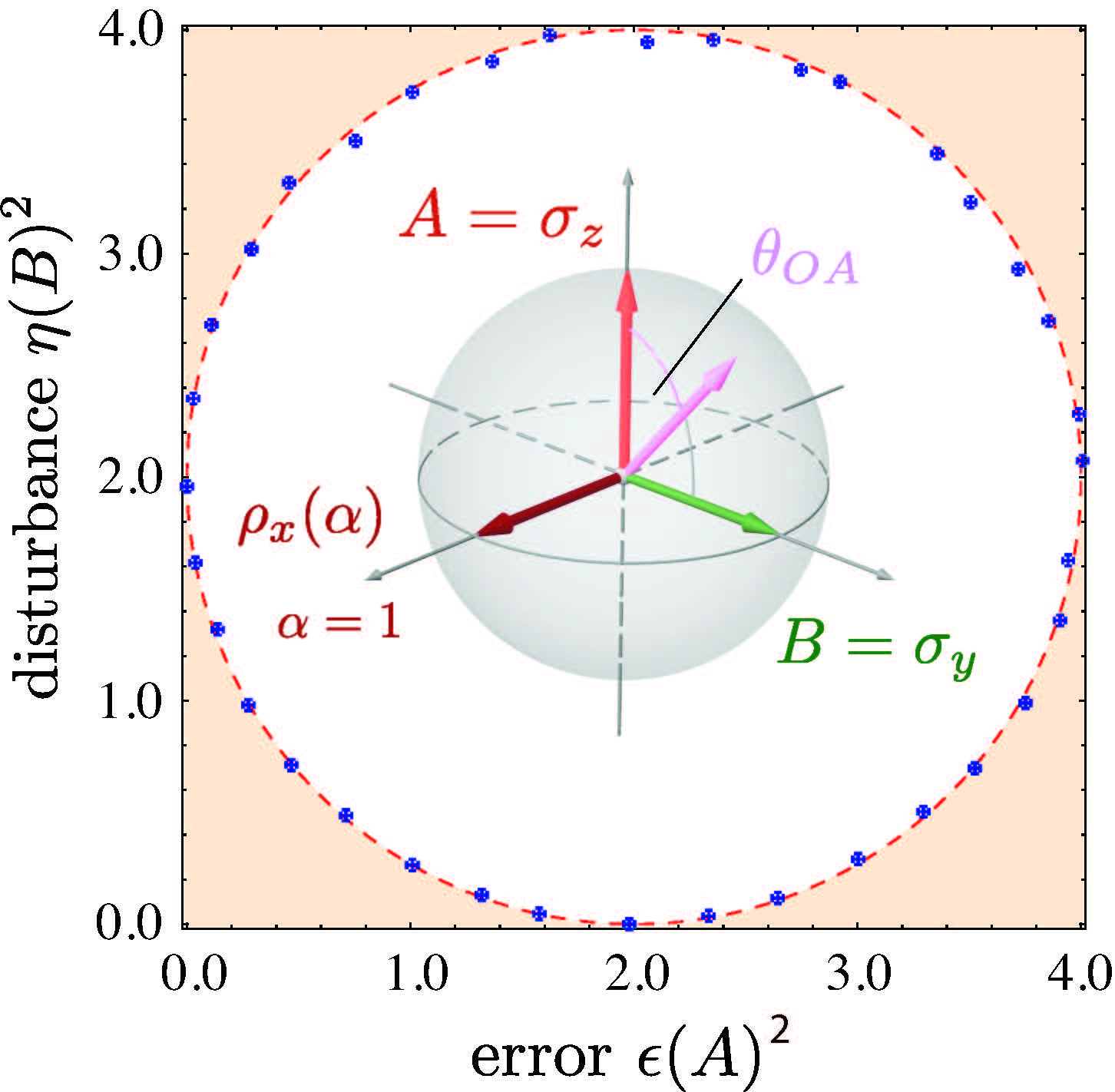}
	\caption{Squared disturbance $\eta(B)^2$ vs. squared error $\epsilon(A)^2$. For detuning angle of $\theta_{OA}=0$ the output observable $O_A$ coincides with $A = \sigma_z$ with $(\epsilon(A), \eta(B)) = (0,2)$. For increasing angles $\theta_{OA}$ the error increases, while disturbance either minimizes when corrected or maximizes when anti-corrected.  }
	\label{fig:errDistSq}
\end{figure}

In Fig.~{\ref{fig:fqd}} typical sets of intensities for different values of $\theta_{OA}$ and $\theta_B$ for five different mixtures $\alpha = \{1, 0.75, 0.5, 0.25, 0 \}$ of the input state $\rho_x(\alpha)=\frac{1}{2}(\mathbbm{1}+\alpha\sigma_x)$ are depicted.
To determine the error $\epsilon(A)$, intensities have to be measured for the input state $\rho=\rho_x(\alpha)$, the auxiliary state $A \rho A=\rho_{-x}(\alpha)$ and  $\rho_{|A} = P^{+}_{A} \rho P^{+}_{A}/\text{Tr}(P^{+}_{A} \rho)$ (this is the pure state $\ket{+z}$). For the disturbance, $\eta(B)$ the input states of  $\rho$, $B \rho B=\rho_{-x}(\alpha)$ and the pure state $\rho_{|B} = P^{+}_{B} \rho P^{+}_{B}/\text{Tr}(P^{+}_{B} \rho)$ are prepared. In Fig.~{\ref{fig:fqd}} $P^{+}_{B} \rho P^{+}_{B}$ is $\ket{+y}$ in (a) and  $\frac{\sqrt 3}{2}\vert +z\rangle+\frac{\rm i}{2}\vert -z\rangle$ in (b), which are the eigenstate of $B$ for $\theta_B=\frac{\pi}{2}$ and $\frac{\pi}{3}$. The pre-factors $\text{Tr}(P^{A}_{+} \rho)$ and $\text{Tr}(P^{B}_{+} \rho)$ are measured separately, by applying only a single apparatus.
The resulting values of the squared error $\epsilon(A)^2$ and the squared disturbance $\eta(B)^2$ are plotted in  Fig.~\ref{fig:errDistSq} for corrected and anti-corrected case, under variation of $\theta_{OA}$.

\section{Mixed state generation}
\label{sec.MixedState}

For a general mixed state given by $\rho_i=\frac{1}{2}(\mathbbm{1}+r_i\sigma_i)$ ($\sigma_i$ represents the direction of the Bloch vector and is given by Pauli operators), the degree of
polarization is given by $P={\rm Tr}(r_i\sigma_i)$.
In order to prepare the mixed states, required for the determination of error $\epsilon(A)$ and disturbance $\eta(B)$, this $P$ has to be varied. This is achieved by applying a random noisy magnetic field in addition to the static one in DC1 (see FIG.~1 in the main manuscript). That is, neutrons with different arriving times
at the coil DC1 experience different magnetic field strengths. This is equivalent to apply different unitary operators $U_{noise}$, describing the noisy $\pi$/2-rotation about the $x$-axes, at each time: this is written in a form of $U_{\rm noise}(\pi/2+\Delta\xi(t):0,-\pi/2)=\widetilde{U}_{\rm noise}(\Delta\xi(t))$ (the terms 0 and -$\pi/2$ denote the polar and azimuthal angle of the rotation axis i.e. the $x$-axis in our case).
For the whole ensemble we have to take the time integral.
Although transformation at each time is unitary, this procedure as a whole ends up as a non-unitary operation due to due to the randomness of the noisy signal and prepares mixed states \cite{Klepp08B}.
For the preparation of the input state $\rho$ (and $A\rho A$), DC1 is positioned in such way that $\rho=\rho_x(\alpha)=\frac{1}{2}(\mathbbm{1}+\alpha\sigma_x)$ is generated at the end of the preparations section, depicted in  FIG.~1 in the main manuscript.


\begin{thebibliography}{34}%
	\makeatletter
	\providecommand \@ifxundefined [1]{%
		\@ifx{#1\undefined}
	}%
	\providecommand \@ifnum [1]{%
		\ifnum #1\expandafter \@firstoftwo
		\else \expandafter \@secondoftwo
		\fi
	}%
	\providecommand \@ifx [1]{%
		\ifx #1\expandafter \@firstoftwo
		\else \expandafter \@secondoftwo
		\fi
	}%
	\providecommand \natexlab [1]{#1}%
	\providecommand \enquote  [1]{``#1''}%
	\providecommand \bibnamefont  [1]{#1}%
	\providecommand \bibfnamefont [1]{#1}%
	\providecommand \citenamefont [1]{#1}%
	\providecommand \href@noop [0]{\@secondoftwo}%
	\providecommand \href [0]{\begingroup \@sanitize@url \@href}%
	\providecommand \@href[1]{\@@startlink{#1}\@@href}%
	\providecommand \@@href[1]{\endgroup#1\@@endlink}%
	\providecommand \@sanitize@url [0]{\catcode `\\12\catcode `\$12\catcode
		`\&12\catcode `\#12\catcode `\^12\catcode `\_12\catcode `\%12\relax}%
	\providecommand \@@startlink[1]{}%
	\providecommand \@@endlink[0]{}%
	\providecommand \url  [0]{\begingroup\@sanitize@url \@url }%
	\providecommand \@url [1]{\endgroup\@href {#1}{\urlprefix }}%
	\providecommand \urlprefix  [0]{URL }%
	\providecommand \doibase [0]{http://dx.doi.org/}%
	\providecommand \selectlanguage [0]{\@gobble}%
	\providecommand \bibinfo  [0]{\@secondoftwo}%
	\providecommand \bibfield  [0]{\@secondoftwo}%
	\providecommand \translation [1]{[#1]}%
	\providecommand \BibitemOpen [0]{}%
	\providecommand \bibitemStop [0]{}%
	\providecommand \bibitemNoStop [0]{.\EOS\space}%
	\providecommand \EOS [0]{\spacefactor3000\relax}%
	\providecommand \BibitemShut  [1]{\csname bibitem#1\endcsname}%
	\let\auto@bib@innerbib\@empty
	\bibitem [{\citenamefont {Wheeler}\ and\ \citenamefont
		{Zurek}(1983)}]{ZurekWheelerBook}%
	\BibitemOpen
	\bibfield  {author} {\bibinfo {author} {\bibfnamefont {J.~A.}\ \bibnamefont
			{Wheeler}}\ and\ \bibinfo {author} {\bibfnamefont {W.~H.}\ \bibnamefont
			{Zurek}},\ }\href@noop {} {\emph {\bibinfo {title} {Quantum Theory and
				Measurement}}}\ (\bibinfo  {publisher} {Princeton Univ. Press, Princeton},\ \bibinfo
	{year} {1983})\BibitemShut {NoStop}%
	\bibitem [{\citenamefont {Born}(1926)}]{Born26}%
	\BibitemOpen
	\bibfield  {author} {\bibinfo {author} {\bibfnamefont {M.}~\bibnamefont
			{Born}},\ }\href {\doibase 10.1007/BF01397477} {\bibfield  {journal}
		{\bibinfo  {journal} {Z. Phys.}\ }\textbf {\bibinfo {volume} {37}},\ \bibinfo
		{pages} {863} (\bibinfo {year} {1926})}\BibitemShut {NoStop}%
	\bibitem [{\citenamefont {Neumann}(1927)}]{Neumann27}%
	\BibitemOpen
	\bibfield  {author} {\bibinfo {author} {\bibfnamefont {J.~v.}\ \bibnamefont
			{Neumann}},\ }{\bibfield  {journal}
		{\bibinfo  {journal} {Mathematische Grundlagen der Quantenmechanik}\ }(\bibinfo {publisher} {Springer, Berlin},\ \bibinfo
		{year} {1932})}\BibitemShut
	{NoStop}%
	\bibitem [{\citenamefont {Sakurai}(1994)}]{Sakurai}%
	\BibitemOpen
	\bibfield  {author} {\bibinfo {author} {\bibfnamefont {J.~J.}\ \bibnamefont
			{Sakurai}},\ }\href@noop {} {\emph {\bibinfo {title} {Modern Quantum
				Mechanics}}}\ (\bibinfo  {publisher} {Addison-Wesley, New York},\ \bibinfo
	{year} {1994})\BibitemShut {NoStop}%
	\bibitem [{\citenamefont {Ballentine}(1998)}]{Ballentine}%
	\BibitemOpen
	\bibfield  {author} {\bibinfo {author} {\bibfnamefont {L.~E.}\ \bibnamefont
			{Ballentine}},\ }\href@noop {} {\emph {\bibinfo {title} {Quantum Mechanics: A
				Modern Development}}}\ (\bibinfo  {publisher} {World Science, New York},\
	\bibinfo {year} {1998})\BibitemShut {NoStop}%
	\bibitem [{\citenamefont {S{\"u}ssmann}(1958)}]{Suesmann58}%
	\BibitemOpen
	\bibfield  {author} {\bibinfo {author} {\bibfnamefont {G.}~\bibnamefont
			{S{\"u}ssmann}},\ }\href
	{http://www.chbeck.de/S%C3%BC%C3%9Fmann-%C3%9Cber-Me%C3%9Fvorgang/productview.aspx?product=17873}
		{\emph {\bibinfo {title} {{\"U}ber den Messvorgang}}},\  (\bibinfo  {publisher} {Abh. Bayer. Akad. Wiss \textbf{88}},\ \bibinfo {year} {1958})\BibitemShut
		{NoStop}%
		\bibitem [{\citenamefont {Englert}(2013)}]{Englert13}%
		\BibitemOpen
		\bibfield  {author} {\bibinfo {author} {\bibfnamefont {B.-G.}\ \bibnamefont
				{Englert}},\ }\href {\doibase 10.1140/epjd/e2013-40486-5} {\bibfield
			{journal} {\bibinfo  {journal} {The European Physical Journal D}\ }\textbf
			{\bibinfo {volume} {67}},\ \bibinfo {eid} {238} (\bibinfo {year} {2013})}\BibitemShut {NoStop}%
	\bibitem [{\citenamefont {Zurek}(2003)}]{Zurek03}%
	\BibitemOpen
	\bibfield  {author} {\bibinfo {author} {\bibfnamefont {W.~H.}\ \bibnamefont
			{Zurek}},\ }\href {\doibase 10.1103/RevModPhys.75.715} {\bibfield  {journal}
		{\bibinfo  {journal} {Rev. Mod. Phys.}\ }\textbf {\bibinfo {volume} {75}},\
		\bibinfo {pages} {715} (\bibinfo {year} {2003})}\BibitemShut {NoStop}%
	\bibitem [{\citenamefont {E.Joos}\ \emph {et~al.}(2003)\citenamefont {E.Joos},
		\citenamefont {H.D.Zeh}, \citenamefont {Kiefer}, \citenamefont {Giulini},
		\citenamefont {Kupsch},\ and\ \citenamefont {Stamatescu}}]{GiuliniBook}%
	\BibitemOpen
	\bibfield  {author} {\bibinfo {author} {\bibnamefont {E.}~\bibnamefont {Joos}}, \bibinfo
		{author} {\bibnamefont {H.D.}~\bibnamefont {Zeh}}, \bibinfo {author} {\bibfnamefont
			{C.}~\bibnamefont {Kiefer}}, \bibinfo {author} {\bibfnamefont
			{D.}~\bibnamefont {Giulini}}, \bibinfo {author} {\bibfnamefont
			{J.}~\bibnamefont {Kupsch}}, \ and\ \bibinfo {author} {\bibfnamefont {I.-O.}\
			\bibnamefont {Stamatescu}},\ }\href
	{http://www.springer.com/us/book/9783540003908} {\emph {\bibinfo {title}
			{Decoherence and the Appearance of a Classical World in Quantum Theory}}}\
	(\bibinfo  {publisher} {Springer, Berlin},\ \bibinfo {year} {2003})\BibitemShut
	{NoStop}%
	\bibitem [{\citenamefont {Heisenberg}(1927)}]{Heisenberg27}%
	\BibitemOpen
	\bibfield  {author} {\bibinfo {author} {\bibfnamefont {W.}~\bibnamefont
			{Heisenberg}},\ }\href {\doibase 10.1007/BF01397280} {\bibfield  {journal}
		{\bibinfo  {journal} {Z. Phys.}\ }\textbf {\bibinfo {volume} {43}},\ \bibinfo
		{pages} {172} (\bibinfo {year} {1927})}\BibitemShut {NoStop}%
		\bibitem [{\citenamefont {Kennard}(1927)}]{Kennard27}%
		\BibitemOpen
		\bibfield  {author} {\bibinfo {author} {\bibfnamefont {E.}~\bibnamefont
				{Kennard}},\ }\href {\doibase 10.1007/BF01391200} {\bibfield  {journal}
			{\bibinfo  {journal} {Z. Phys.}\ }\textbf {\bibinfo {volume} {44}},\ \bibinfo
			{pages} {326} (\bibinfo {year} {1927})}\BibitemShut {NoStop}%
		\bibitem [{\citenamefont {Robertson}(1929)}]{Robertson29}%
		\BibitemOpen
		\bibfield  {author} {\bibinfo {author} {\bibfnamefont {H.~P.}\ \bibnamefont
				{Robertson}},\ }\href {\doibase 10.1103/PhysRev.34.163} {\bibfield  {journal}
			{\bibinfo  {journal} {Phys. Rev.}\ }\textbf {\bibinfo {volume} {34}},\
			\bibinfo {pages} {163} (\bibinfo {year} {1929})}\BibitemShut {NoStop}%
	\bibitem [{\citenamefont {Ozawa}(2003{\natexlab{a}})}]{Ozawa03}%
	\BibitemOpen
	\bibfield  {author} {\bibinfo {author} {\bibfnamefont {M.}~\bibnamefont
			{Ozawa}},\ }\href {\doibase 10.1103/PhysRevA.67.042105} {\bibfield  {journal}
		{\bibinfo  {journal} {Phys. Rev. A}\ }\textbf {\bibinfo {volume} {67}},\
		\bibinfo {pages} {042105} (\bibinfo {year} {2003}{\natexlab{a}})}\BibitemShut
	{NoStop}%
	\bibitem [{\citenamefont {Ozawa}(2004)}]{Ozawa04}%
	\BibitemOpen
	\bibfield  {author} {\bibinfo {author} {\bibfnamefont {M.}~\bibnamefont
			{Ozawa}},\ }\href
	{http://www.sciencedirect.com/science/article/pii/S0003491604000089}
	{\bibfield  {journal} {\bibinfo  {journal} {Annals of Physics}\ }\textbf
		{\bibinfo {volume} {311}},\ \bibinfo {pages} {350 } (\bibinfo {year}
		{2004})}\BibitemShut {NoStop}%
	\bibitem [{\citenamefont {Erhart}\ \emph {et~al.}(2012)\citenamefont {Erhart},
		\citenamefont {Sponar}, \citenamefont {Sulyok}, \citenamefont {Badurek},
		\citenamefont {Ozawa},\ and\ \citenamefont {Hasegawa}}]{Erhart12}%
	\BibitemOpen
	\bibfield  {author} {\bibinfo {author} {\bibfnamefont {J.}~\bibnamefont
			{Erhart}}, \bibinfo {author} {\bibfnamefont {S.}~\bibnamefont {Sponar}},
		\bibinfo {author} {\bibfnamefont {G.}~\bibnamefont {Sulyok}}, \bibinfo
		{author} {\bibfnamefont {G.}~\bibnamefont {Badurek}}, \bibinfo {author}
		{\bibfnamefont {M.}~\bibnamefont {Ozawa}}, \ and\ \bibinfo {author}
		{\bibfnamefont {Y.}~\bibnamefont {Hasegawa}},\ }\href
	{http://dx.doi.org/10.1038/nphys2194} {\bibfield  {journal} {\bibinfo
			{journal} {Nat Phys}\ }\textbf {\bibinfo {volume} {8}},\ \bibinfo {pages}
		{185} (\bibinfo {year} {2012})}\BibitemShut {NoStop}%
	\bibitem [{\citenamefont {Sulyok}\ \emph {et~al.}(2013)\citenamefont {Sulyok},
		\citenamefont {Sponar}, \citenamefont {Erhart}, \citenamefont {Badurek},
		\citenamefont {Ozawa},\ and\ \citenamefont {Hasegawa}}]{Sulyok13}%
	\BibitemOpen
	\bibfield  {author} {\bibinfo {author} {\bibfnamefont {G.}~\bibnamefont
			{Sulyok}}, \bibinfo {author} {\bibfnamefont {S.}~\bibnamefont {Sponar}},
		\bibinfo {author} {\bibfnamefont {J.}~\bibnamefont {Erhart}}, \bibinfo
		{author} {\bibfnamefont {G.}~\bibnamefont {Badurek}}, \bibinfo {author}
		{\bibfnamefont {M.}~\bibnamefont {Ozawa}}, \ and\ \bibinfo {author}
		{\bibfnamefont {Y.}~\bibnamefont {Hasegawa}},\ }\href {\doibase
		10.1103/PhysRevA.88.022110} {\bibfield  {journal} {\bibinfo  {journal} {Phys.
				Rev. A}\ }\textbf {\bibinfo {volume} {88}},\ \bibinfo {pages} {022110}
		(\bibinfo {year} {2013})}\BibitemShut {NoStop}%
	\bibitem [{\citenamefont {Rozema}\ \emph {et~al.}(2012)\citenamefont {Rozema},
		\citenamefont {Darabi}, \citenamefont {Mahler}, \citenamefont {Hayat},
		\citenamefont {Soudagar},\ and\ \citenamefont {Steinberg}}]{Steinberg12}%
	\BibitemOpen
	\bibfield  {author} {\bibinfo {author} {\bibfnamefont {L.~A.}\ \bibnamefont
			{Rozema}}, \bibinfo {author} {\bibfnamefont {A.}~\bibnamefont {Darabi}},
		\bibinfo {author} {\bibfnamefont {D.~H.}\ \bibnamefont {Mahler}}, \bibinfo
		{author} {\bibfnamefont {A.}~\bibnamefont {Hayat}}, \bibinfo {author}
		{\bibfnamefont {Y.}~\bibnamefont {Soudagar}}, \ and\ \bibinfo {author}
		{\bibfnamefont {A.~M.}\ \bibnamefont {Steinberg}},\ }\href {\doibase
		10.1103/PhysRevLett.109.100404} {\bibfield  {journal} {\bibinfo  {journal}
			{Phys. Rev. Lett.}\ }\textbf {\bibinfo {volume} {109}},\ \bibinfo {pages}
		{100404} (\bibinfo {year} {2012})}\BibitemShut {NoStop}%
	\bibitem [{\citenamefont {Baek}\ \emph {et~al.}(2013)\citenamefont {Baek},
		\citenamefont {Kaneda}, \citenamefont {Ozawa},\ and\ \citenamefont
		{Edamatsu}}]{Edamatsu13}%
	\BibitemOpen
	\bibfield  {author} {\bibinfo {author} {\bibfnamefont {S.-Y.}\ \bibnamefont
			{Baek}}, \bibinfo {author} {\bibfnamefont {F.}~\bibnamefont {Kaneda}},
		\bibinfo {author} {\bibfnamefont {M.}~\bibnamefont {Ozawa}}, \ and\ \bibinfo
		{author} {\bibfnamefont {K.}~\bibnamefont {Edamatsu}},\ }\href
	{http://www.nature.com/srep/2013/130717/srep02221/abs/srep02221.html\#supplementary-information}
	{\bibfield  {journal} {\bibinfo  {journal} {Sci. Rep.}\ }\textbf {\bibinfo
			{volume} {3}} (\bibinfo {year} {2013})}\BibitemShut {NoStop}%
	\bibitem [{\citenamefont {Weston}\ \emph {et~al.}(2013)\citenamefont {Weston},
		\citenamefont {Hall}, \citenamefont {Palsson}, \citenamefont {Wiseman},\ and\
		\citenamefont {Pryde}}]{Weston13}%
	\BibitemOpen
	\bibfield  {author} {\bibinfo {author} {\bibfnamefont {M.~M.}\ \bibnamefont
			{Weston}}, \bibinfo {author} {\bibfnamefont {M.~J.~W.}\ \bibnamefont {Hall}},
		\bibinfo {author} {\bibfnamefont {M.~S.}\ \bibnamefont {Palsson}}, \bibinfo
		{author} {\bibfnamefont {H.~M.}\ \bibnamefont {Wiseman}}, \ and\ \bibinfo
		{author} {\bibfnamefont {G.~J.}\ \bibnamefont {Pryde}},\ }\href {\doibase
		10.1103/PhysRevLett.110.220402} {\bibfield  {journal} {\bibinfo  {journal}
			{Phys. Rev. Lett.}\ }\textbf {\bibinfo {volume} {110}},\ \bibinfo {pages}
		{220402} (\bibinfo {year} {2013})}\BibitemShut {NoStop}%
		\bibitem [{\citenamefont {Busch}\ \emph {et~al.}(2013)\citenamefont {Busch},
			\citenamefont {Lahti},\ and\ \citenamefont {Werner}}]{Busch13}%
		\BibitemOpen
		\bibfield  {author} {\bibinfo {author} {\bibfnamefont {P.}~\bibnamefont
				{Busch}}, \bibinfo {author} {\bibfnamefont {P.}~\bibnamefont {Lahti}}, \ and\
			\bibinfo {author} {\bibfnamefont {R.~F.}\ \bibnamefont {Werner}},\ }\href
		{\doibase 10.1103/PhysRevLett.111.160405} {\bibfield  {journal} {\bibinfo
				{journal} {Phys. Rev. Lett.}\ }\textbf {\bibinfo {volume} {111}},\ \bibinfo
			{pages} {160405} (\bibinfo {year} {2013})}\BibitemShut {NoStop}%
		\bibitem [{\citenamefont {Busch}\ \emph {et~al.}(2013)\citenamefont {Busch},
			\citenamefont {Lahti},\ and\ \citenamefont {Werner}}]{Busch14}%
		\BibitemOpen
		\bibfield  {author} {\bibinfo {author} {\bibfnamefont {P.}~\bibnamefont
				{Busch}}, \bibinfo {author} {\bibfnamefont {P.}~\bibnamefont {Lahti}}, \ and\
			\bibinfo {author} {\bibfnamefont {R.~F.}\ \bibnamefont {Werner}},\ }\href
		{\doibase 10.1103/RevModPhys.86.1261} {\bibfield  {journal} {\bibinfo
				{journal} {Rev. Mod. Phys.}\ }\textbf {\bibinfo {volume} {86}},\ \bibinfo
			{pages} {1261} (\bibinfo {year} {2014})}\BibitemShut {NoStop}%
		\bibitem [{\citenamefont {Buscemi}\ \emph {et~al.}(2014)\citenamefont
			{Buscemi}, \citenamefont {Hall}, \citenamefont {Ozawa},\ and\ \citenamefont
			{Wilde}}]{Buscemi14}%
		\BibitemOpen
		\bibfield  {author} {\bibinfo {author} {\bibfnamefont {F.}~\bibnamefont
				{Buscemi}}, \bibinfo {author} {\bibfnamefont {M.~J.~W.}\ \bibnamefont
				{Hall}}, \bibinfo {author} {\bibfnamefont {M.}~\bibnamefont {Ozawa}}, \ and\
			\bibinfo {author} {\bibfnamefont {M.~M.}\ \bibnamefont {Wilde}},\ }\href
		{\doibase 10.1103/PhysRevLett.112.050401} {\bibfield  {journal} {\bibinfo
				{journal} {Phys. Rev. Lett.}\ }\textbf {\bibinfo {volume} {112}},\ \bibinfo
			{pages} {050401} (\bibinfo {year} {2014})}\BibitemShut {NoStop}%
		\bibitem [{\citenamefont {Lu}\ \emph {et~al.}(2014)\citenamefont {Lu},
			\citenamefont {Yu}, \citenamefont {Fujikawa},\ and\ \citenamefont
			{Oh}}]{Lu14}%
		\BibitemOpen
		\bibfield  {author} {\bibinfo {author} {\bibfnamefont {X.-M.}\ \bibnamefont
				{Lu}}, \bibinfo {author} {\bibfnamefont {S.}~\bibnamefont {Yu}}, \bibinfo
			{author} {\bibfnamefont {K.}~\bibnamefont {Fujikawa}}, \ and\ \bibinfo
			{author} {\bibfnamefont {C.~H.}\ \bibnamefont {Oh}},\ }\href {\doibase
			10.1103/PhysRevA.90.042113} {\bibfield  {journal} {\bibinfo  {journal} {Phys.
					Rev. A}\ }\textbf {\bibinfo {volume} {90}},\ \bibinfo {pages} {042113}
			(\bibinfo {year} {2014})}\BibitemShut {NoStop}%
	\bibitem [{\citenamefont {Branciard}(2013)}]{Branciard13}%
	\BibitemOpen
	\bibfield  {author} {\bibinfo {author} {\bibfnamefont {C.}~\bibnamefont
			{Branciard}},\ }\href {\doibase 10.1073/pnas.1219331110} {\bibfield
		{journal} {\bibinfo  {journal} {Proc. Natl. Acad. Sci. USA}\ }\textbf
		{\bibinfo {volume} {17}},\ \bibinfo {pages} {6742} (\bibinfo {year}
		{2013})}\BibitemShut {NoStop}%
		\bibitem [{\citenamefont {Branciard}(2014)}]{Branciard14}%
		\BibitemOpen
		\bibfield  {author} {\bibinfo {author} {\bibfnamefont {C.}~\bibnamefont
				{Branciard}},\ }\href {\doibase 10.1103/PhysRevA.89.022124} {\bibfield
			{journal} {\bibinfo  {journal} {Phys. Rev. A}\ }\textbf {\bibinfo {volume}
				{89}},\ \bibinfo {pages} {022124} (\bibinfo {year} {2014})}\BibitemShut
		{NoStop}%
		\bibitem [{\citenamefont {Ozawa}(2014)}]{Ozawa14}%
		\BibitemOpen
		\bibfield  {author} {\bibinfo {author} {\bibfnamefont {M.}~\bibnamefont
				{Ozawa}},\ }\href {http://arxiv.org/abs/1404.3388} {\bibfield  {journal}
			{\bibinfo  {journal} {arXiv:1404.3388v1 [quant-ph]}\ } (\bibinfo {year}
			{2014})}\BibitemShut {NoStop}%
		\bibitem [{\citenamefont {Sponar}\ \emph {et~al.}(2015)\citenamefont {Sponar},
			\citenamefont {Sulyok}, \citenamefont {Erhart},\ and\ \citenamefont
			{Hasegawa}}]{Sponar14}%
		\BibitemOpen
		\bibfield  {author} {\bibinfo {author} {\bibfnamefont {S.}~\bibnamefont
				{Sponar}}, \bibinfo {author} {\bibfnamefont {G.}~\bibnamefont {Sulyok}},
			\bibinfo {author} {\bibfnamefont {J.}~\bibnamefont {Erhart}}, \ and\ \bibinfo
			{author} {\bibfnamefont {Y.}~\bibnamefont {Hasegawa}},\ }\href {\doibase
			10.1155/2014/735398} {\bibfield  {journal} {\bibinfo  {journal} {Adv. High
					Energy Phys.}\ }\textbf {\bibinfo {volume} {44}},\ \bibinfo {pages} {36}
			(\bibinfo {year} {2015})}\BibitemShut {NoStop}%
		\bibitem [{\citenamefont {Klepp}\ \emph {et~al.}(2008)\citenamefont {Klepp},
			\citenamefont {Sponar}, \citenamefont {Filipp}, \citenamefont {Lettner},
			\citenamefont {Badurek},\ and\ \citenamefont {Hasegawa}}]{Klepp08}%
		\BibitemOpen
		\bibfield  {author} {\bibinfo {author} {\bibfnamefont {J.}~\bibnamefont
				{Klepp}}, \bibinfo {author} {\bibfnamefont {S.}~\bibnamefont {Sponar}},
			\bibinfo {author} {\bibfnamefont {S.}~\bibnamefont {Filipp}}, \bibinfo
			{author} {\bibfnamefont {M.}~\bibnamefont {Lettner}}, \bibinfo {author}
			{\bibfnamefont {G.}~\bibnamefont {Badurek}}, \ and\ \bibinfo {author}
			{\bibfnamefont {Y.}~\bibnamefont {Hasegawa}},\ }\href {\doibase
			10.1103/PhysRevLett.101.150404} {\bibfield  {journal} {\bibinfo  {journal}
				{Phys. Rev. Lett.}\ }\textbf {\bibinfo {volume} {101}},\ \bibinfo {pages}
			{150404} (\bibinfo {year} {2008})}\BibitemShut {NoStop}%
	\bibitem [{\citenamefont {Ringbauer}\ \emph {et~al.}(2014)\citenamefont
		{Ringbauer}, \citenamefont {Biggerstaff}, \citenamefont {Broome},
		\citenamefont {Fedrizzi}, \citenamefont {Branciard},\ and\ \citenamefont
		{White}}]{Ringbauer14}%
	\BibitemOpen
	\bibfield  {author} {\bibinfo {author} {\bibfnamefont {M.}~\bibnamefont
			{Ringbauer}}, \bibinfo {author} {\bibfnamefont {D.~N.}\ \bibnamefont
			{Biggerstaff}}, \bibinfo {author} {\bibfnamefont {M.~A.}\ \bibnamefont
			{Broome}}, \bibinfo {author} {\bibfnamefont {A.}~\bibnamefont {Fedrizzi}},
		\bibinfo {author} {\bibfnamefont {C.}~\bibnamefont {Branciard}}, \ and\
		\bibinfo {author} {\bibfnamefont {A.~G.}\ \bibnamefont {White}},\ }\href
	{\doibase 10.1103/PhysRevLett.112.020401} {\bibfield  {journal} {\bibinfo
			{journal} {Phys. Rev. Lett.}\ }\textbf {\bibinfo {volume} {112}},\ \bibinfo
		{pages} {020401} (\bibinfo {year} {2014})}\BibitemShut {NoStop}%
	\bibitem [{\citenamefont {Kaneda}\ \emph {et~al.}(2014)\citenamefont {Kaneda},
		\citenamefont {Baek}, \citenamefont {Ozawa},\ and\ \citenamefont
		{Edamatsu}}]{Kaneda14}%
	\BibitemOpen
	\bibfield  {author} {\bibinfo {author} {\bibfnamefont {F.}~\bibnamefont
			{Kaneda}}, \bibinfo {author} {\bibfnamefont {S.-Y.}\ \bibnamefont {Baek}},
		\bibinfo {author} {\bibfnamefont {M.}~\bibnamefont {Ozawa}}, \ and\ \bibinfo
		{author} {\bibfnamefont {K.}~\bibnamefont {Edamatsu}},\ }\href {\doibase
		10.1103/PhysRevLett.112.020402} {\bibfield  {journal} {\bibinfo  {journal}
			{Phys. Rev. Lett.}\ }\textbf {\bibinfo {volume} {112}},\ \bibinfo {pages}
		{020402} (\bibinfo {year} {2014})}\BibitemShut {NoStop}%
		\bibitem [{\citenamefont {Kirchmair}\ \emph {et~al.}(2009)\citenamefont
			{Kirchmair}, \citenamefont {Z{\"a}hringer}, \citenamefont {Gerritsma},
			\citenamefont {Kleinmann}, \citenamefont {Guhne}, \citenamefont {Cabello},
			\citenamefont {Blatt},\ and\ \citenamefont {Roos}}]{Kirchmair2009}%
		\BibitemOpen
		\bibfield  {author} {\bibinfo {author} {\bibfnamefont {G.}~\bibnamefont
				{Kirchmair}}, \bibinfo {author} {\bibfnamefont {F.}~\bibnamefont
				{Z{\"a}hringer}}, \bibinfo {author} {\bibfnamefont {R.}~\bibnamefont
				{Gerritsma}}, \bibinfo {author} {\bibfnamefont {M.}~\bibnamefont
				{Kleinmann}}, \bibinfo {author} {\bibfnamefont {O.}~\bibnamefont {Guhne}},
			\bibinfo {author} {\bibfnamefont {A.}~\bibnamefont {Cabello}}, \bibinfo
			{author} {\bibfnamefont {R.}~\bibnamefont {Blatt}}, \ and\ \bibinfo {author}
			{\bibfnamefont {C.~F.}\ \bibnamefont {Roos}},\ }\href
		{http://dx.doi.org/10.1038/nature08172
			http://www.nature.com/nature/journal/v460/n7254/suppinfo/nature08172\_S1.html}
		{\bibfield  {journal} {\bibinfo  {journal} {Nature}\ }\textbf {\bibinfo
				{volume} {460}},\ \bibinfo {pages} {494} (\bibinfo {year}
			{2009})}\BibitemShut {NoStop}%
	\end{thebibliography}

\begin{thebibliography}{2}
	\expandafter\ifx\csname natexlab\endcsname\relax\def\natexlab#1{#1}\fi
	\expandafter\ifx\csname bibnamefont\endcsname\relax
	\def\bibnamefont#1{#1}\fi
	\expandafter\ifx\csname bibfnamefont\endcsname\relax
	\def\bibfnamefont#1{#1}\fi
	\expandafter\ifx\csname citenamefont\endcsname\relax
	\def\citenamefont#1{#1}\fi
	\expandafter\ifx\csname url\endcsname\relax
	\def\url#1{\texttt{#1}}\fi
	\expandafter\ifx\csname urlprefix\endcsname\relax\def\urlprefix{URL }\fi
	\providecommand{\bibinfo}[2]{#2}
	\providecommand{\eprint}[2][]{\url{#2}}
	
	\bibitem [{\citenamefont {Ozawa}(2004)}]{Ozawa04B}%
	\BibitemOpen
	\bibfield  {author} {\bibinfo {author} {\bibfnamefont {M.}~\bibnamefont
			{Ozawa}},\ }\href
	{http://www.sciencedirect.com/science/article/pii/S0003491604000089}
	{\bibfield  {journal} {\bibinfo  {journal} {Annals of Physics}\ }\textbf
		{\bibinfo {volume} {311}},\ \bibinfo {pages} {350 } (\bibinfo {year}
		{2004})}\BibitemShut {NoStop}%
	\bibitem [{\citenamefont {Ozawa}(2003{\natexlab{a}})}]{Ozawa03B}%
	\BibitemOpen
	\bibfield  {author} {\bibinfo {author} {\bibfnamefont {M.}~\bibnamefont
			{Ozawa}},\ }\href {\doibase 10.1103/PhysRevA.67.042105} {\bibfield  {journal}
		{\bibinfo  {journal} {Phys. Rev. A}\ }\textbf {\bibinfo {volume} {67}},\
		\bibinfo {pages} {042105} (\bibinfo {year} {2003}{\natexlab{a}})}\BibitemShut
	{NoStop}%
	\bibitem [{\citenamefont {Branciard}(2013)}]{Branciard13B}%
	\BibitemOpen
	\bibfield  {author} {\bibinfo {author} {\bibfnamefont {C.}~\bibnamefont
			{Branciard}},\ }\href {\doibase 10.1073/pnas.1219331110} {\bibfield
		{journal} {\bibinfo  {journal} {Proc. Natl. Acad. Sci. USA}\ }\textbf
		{\bibinfo {volume} {17}},\ \bibinfo {pages} {6742} (\bibinfo {year}
		{2013})}\BibitemShut {NoStop}%
	\bibitem [{\citenamefont {Klepp}\ \emph {et~al.}(2008)\citenamefont {Klepp},
		\citenamefont {Sponar}, \citenamefont {Filipp}, \citenamefont {Lettner},
		\citenamefont {Badurek},\ and\ \citenamefont {Hasegawa}}]{Klepp08B}%
	\BibitemOpen
	\bibfield  {author} {\bibinfo {author} {\bibfnamefont {J.}~\bibnamefont
			{Klepp}}, \bibinfo {author} {\bibfnamefont {S.}~\bibnamefont {Sponar}},
		\bibinfo {author} {\bibfnamefont {S.}~\bibnamefont {Filipp}}, \bibinfo
		{author} {\bibfnamefont {M.}~\bibnamefont {Lettner}}, \bibinfo {author}
		{\bibfnamefont {G.}~\bibnamefont {Badurek}}, \ and\ \bibinfo {author}
		{\bibfnamefont {Y.}~\bibnamefont {Hasegawa}},\ }\href {\doibase
		10.1103/PhysRevLett.101.150404} {\bibfield  {journal} {\bibinfo  {journal}
			{Phys. Rev. Lett.}\ }\textbf {\bibinfo {volume} {101}},\ \bibinfo {pages}
		{150404} (\bibinfo {year} {2008})}\BibitemShut {NoStop}%
\end{thebibliography}
\end{document}